\documentclass[a4paper, 11pt]{article}
\usepackage[a4paper, margin = 3cm]{geometry}

\usepackage[style=authoryear, dashed=false, url=false, doi=true, date=year, isbn=false, sorting=nyt, maxcitenames=2, maxbibnames=30, giveninits=true]{biblatex}

\usepackage[colorlinks=true, citecolor=blue, linkcolor=black, urlcolor=blue]{hyperref}
\usepackage{booktabs}
\usepackage{lipsum}
\usepackage{lineno}
\usepackage{authblk}
\usepackage[autostyle]{csquotes}  
\usepackage{amsmath}
\usepackage{amssymb}
\usepackage{stmaryrd} 
\usepackage{bm}
\usepackage{enumitem}
\usepackage[font=footnotesize]{caption}
\usepackage{graphicx, wrapfig, rotating}

\renewbibmacro{in:}{}

\DeclareFieldFormat{pages}{#1}
\DeclareFieldFormat{postnote}{#1}
\DeclareFieldFormat{multipostnote}{#1}

\providecommand{\keywords}[1]{\textbf{Keywords:} #1}

\begin{document}

\pagenumbering{arabic}
\date{}
\title{Mind the hubris: complexity can misfire}

\author[1,2]{Arnald Puy}
\author[2,3]{Andrea Saltelli}

\affil[1]{\footnotesize{\textit{School of Geography, Earth and Environmental Sciences, University of Birmingham, Birmingham B15 2TT, United Kingdom. E-mail: \href{mailto:a.puy@bham.ac.uk}{a.puy@bham.ac.uk}}}}

\affil[2]{\footnotesize{\textit{Centre for the Study of the Sciences and the Humanities (SVT), University of Bergen, Parkveien 9, PB 7805, 5020 Bergen, Norway.}}}

\affil[3]{\footnotesize{\textit{Barcelona School of Management, Pompeu Fabra University, Carrer de Balmes 132, 08008 Barcelona, Spain.}}}

%\linenumbers

\maketitle

\textbf{This is a draft of a chapter that has been accepted for publication by Oxford University Press in the forthcoming book \textit{Views on Mathematical Modelling}, edited by Andrea Saltelli and Monica di Fiore and due for publication in 2023.}

\begin{abstract}
Here we briefly reflect on the philosophical foundations that ground the quest towards ever-detailed models and identify four practical dangers derived from this pursuit: explosion of the model's uncertainty space, model black-boxing, computational exhaustion and model attachment. We argue that the growth of a mathematical model should be carefully and continuously pondered lest models become extraneous constructs chasing the Cartesian dream.

\end{abstract}


\newpage

\section*{Mathematics and tales}

Possibly the most famous quote of the French philosopher \textcite[51]{Descartes2006} is that of man as ``master and possessor of nature'', an individual that uses the power of mathematical reason to decode the natural order and improve human condition. Technically, Descartes saw more clearly than others how applying algebra to solve geometrical problems would open the door to the solution of practical challenges. This vision was received by Descartes in a dream and has fueled the extraordinary development that science and technology has experienced over the last three centuries. In fact, Descartes' dream has been so successful that ``it would not be a mistake to call our age and all its scientific aspirations Cartesian''~\parencite[260]{Davis1986}.

Descartes' vision of a mechanical world dominated by mathematical laws prone to be deciphered through scientific inquiry was also shared by philosophers such as Galilei, Leibniz, Laplace or Condorcet~\parencite{Funtowicz2020}. Underlying this premise there is the platonic idea of a mathematical truth existing ``out there'' that provides structure to the universe independently of the individual and its cognitive abilities. Some significant defenders of this view in the 19th--20th century were~\textcite{Frege2007} and~\textcite{Godel1951}, who argued against mathematics being a product of the human psyche and endorsed the platonistic view as the ``only one tenable''. Years later, Benacerraf inspired one of the most famous epistemological objections to mathematical platonism by arguing that platonists cannot explain mathematical beliefs because mathematical objects are abstract and hence not causally active --nothing connects the knowledge holder with the object known~\parencite{Benacerraf1973, Field1989}. For~\textcite[176]{Balaguer1998}, the debate between mathematical platonists and anti-platonists may not be solvable given that both sides hold convincing arguments and are perfectly workable philosophies of mathematics.

And yet the assumption that mathematical entities objectively exist for us to appraise through scientific inquiry seems to have sidelined the alternative in the field of mathematical modeling. This is apparent in the trend towards ever-complex models characterizing natural science fields such as hydrology or climate studies: models get bigger to accommodate the newly acquired knowledge on previously hidden mechanisms that had been brought to light by scientific methods and state-of-the-art technologies\footnote{This is especially apparent if one observes the evolution of climate, hydrological and epidemiological models over the last 80 years. For instance, the simple general circulation models of the 1960's have become exhaustive atmosphere-ocean general circulation models~\parencite{Hausfather2020, Sarofim2021}. The classic bucket-type models of the 1970's have turned into global hydrological models that simulate the whole water cycle and the impact of humans in it~\parencite{Manabe1969, Bierkens2015}. In epidemiology, the first compartment models in the 1930s--1940s had just a few parameters~\parencite{Kermack1933}; the covid-19 model of the Imperial College London had  c.~900~\parencite{Ferguson2020}. See~\textcite{Puyn} for further details.}. Uncertainties are regarded as mostly epistemic and prone to be overcome or significantly abated once we dig deeper into the mathematical structure underlying the process under study\footnote{The design of finer-grained models is regarded as a critical step towards that aim. In climate prediction, more detailed models are assumed to solve convective cloud systems and their influence upon the reflection of incoming solar radiation~\parencite{Palmer2014, Schar2020, Fuhrer2018}. In hydrology, finer-grained detail is thought to provide better representations of human water demands, land-atmospheric interactions, topography and vegetation and soil moisture and evapotranspiration processes~\parencite{Bierkens2015, Wood2011}. See~\textcite{Beven2012a, Beven2015} for a critique of the drive towards model hyper-resolution in hydrology.}. Unless it embraces the idea that humans are capable to eventually decipher the main mathematical intricacies of the universe, this current implicitly assumes that the quest for larger and more descriptive models may not have an end.

If an anti-platonic understanding of mathematics is equally defensible, however, we must concede that our mathematical representations of physical phenomena may simply be metaphors without any objective proof. The success of mathematics in describing these phenomena does not necessarily demonstrate its existence\footnote{The assumption that there is a mathematical truth because mathematics is essential to explain the success of science is known as the Quine-Putnam indispensability argument for mathematical realism~\parencite{Putnam1975}, and has been undermined among others by~\textcite{Maddy1992} and \textcite{Sober1993}.}:~\textcite{Field2016} and \textcite{Balaguer1996} showed that Newtonian physics and quantum mechanics can be precisely nominalised and described without mathematics, pointing towards the scientific usefulness of mathematics as being mostly pragmatic, not idealistic --mathematics just makes calculations simpler. Whatever the fit there is between a mathematical representation of a regularity and the regularity itself may hence occur in the mind of the observer and not in the real world. According to~\textcite[81]{Lakoff2000}, this makes mathematics a human (and not divine) phenomenon\footnote{The application of mathematics to the social sciences is also the subject of concern, both for its possible dangerous styles of thinking imposed on the social issue under study and for the performative, non-neutral role of mathematics as an element of education and regimentation~\parencite{Ernest2018}. The mathematization of economics is lamented by~\textcite{Drechsler2000, Mirowski1989, Reinert2019, Romer2015}. See also Drechsler in this volume for a critique of a ``quantitative-mathematical social science''.}:

\begin{quote}
(...) it follows from the empirical study of numbers as a product of mind that it is natural for people to believe that numbers are not a product of mind!
\end{quote}

Mathematical models in this current would be meta-metaphors given their higher-order abstraction of a set of already abstract concepts. By assuming the absence of transcendent mathematics, this view ultimately regards models as figurative representations of an allegory and thus as intrinsically uncertain constructs --their design, scope and conceptualization is not defined by the underlying set of mechanisms modeled  as much as by the subjective perspective of the modeler(s), which inevitably reflects their methodological and disciplinary configuration~\parencite{Saltelli2020}. The addition of model detail as a mean to get closer to the ``truth'' may lead a modeler astray in the pursuit of a ``truth'' that only exists as such in the modeler's mind.

The predominance of the platonic view in mathematical modelling is a distinctive tenet of modernity and its willingness to unfold universal, timeless physical truths over local, timely principles --what~\textcite{Toulmin1990} refers to as the triumph of Descartes' certainty over Montaigne's doubt. The progressive formalization of reality into numbers and equations of increasing sophistication reflects the thirst for developing more objective, rational solutions to the social-environmental challenges that our society faces. A conspicuous example of this ethos is the Destination Earth Initiative launched in 2022 by the~\textcite{EU2022}, which aims at creating a digital twin of the Earth (``a highly accurate model'') to ``monitor, model and predict natural and human activity'', ultimately helping to ``tackle climate change and protect nature''. The same can be said of the willingness to simulate global water flows at a 1 km resolution and every 1-3 h by a section of the hydrological community in order to ``adequately address critical water science questions''~\parencite{Wood2011}. This ambition resembles~\textcite{Carroll1893}'s fictional tale of a nation that developed a map with a scale of a mile to the mile, or~\textcite{Borges1998}' story of the map of the Empire whose ``size was that of the Empire, and which coincided point for point with it''. In both tales, the map ends up being discarded as useless.

Carrol's and Borges' tales do not differ much from what models are in the anti-platonic view of mathematics given their abstraction of ideas with a likeness to the empirical world. And yet their figurative meaning does not diminish their capacity to produce reflection and wisdom: we all get what the insight is and how the fictional stories connect with our experiences. The tales are not ``out there'' but they convey an understanding about a complex, specific feature of human behavior in a plain and simple way. The addition of narrative detail may or may not nuance the story and hence it is not an end in itself. If models can philosophically be thought of as meta-metaphors in a legitimate way, then their metaphysical status may be indistinguishable from that of any literary work --mathematical units such as $\pi$ would be indistinguishable from fictional characters such as Sherlock Holmes~\parencite[345]{Bonevac2009}.

The philosophical validity of the anti-platonic view of mathematics opens the door to important questions with regards to our overreliance on the use of models as tools for prediction, management and control. It also suggests caution in the quest towards ever-complex models so as to achieve sharper insights. Our use of the Greek word ``hubris'' in the title denotes the self-assured arrogance that comes with the lack of attention to these epistemological issues. For~\textcite[238]{Jasanoff2003}, mathematical models are ``technologies of hubris'' when they aim at predicting phenomena to facilitate management and control in areas of high uncertainty. The addition of conceptual depth to a model with unexplored ambiguities makes the hubris problem worse. In the next pages we discuss four specific issues derived from this problem: explosion of the uncertainty space, blackboxing, computational exhaustion and model attachment. We conclude by offering some reflections on why mathematical models may benefit from relaxing the pursuit of the Cartesian dream. 

\section*{Explosion of the uncertainty space}

In the environmental, climate or epidemiological sciences the addition of model detail often involves increasing the spatial/temporal resolution of the model, adding new parameters and feedbacks or enlarging the model with new compartments linked in a causal chain: for instance, a model simulating how an aquifer gets polluted may be extended with compartments that model how this pollution reaches the surface, how it is metabolized by crops and finally how it impacts the health of the affected population~\parencite{Goodwin1987}. During this model expansion process new parameters need to be estimated with their error due to natural variation, measurement bias or disagreement amongst experts about their ``true'' value. There might also be ambiguity as to which is the best way to mathematically integrate the new parameters into the model. All this often makes the model output uncertainty increase, and not decrease, at each model upgrade stage. Scholars refer to this phenomenon as the ``uncertainty cascade'' effect, where uncertainties add up and expand with model complexification~\parencite{Hillerbrand2014, Wilby2010}. 

\begin{figure}[ht]
\centering
\includegraphics[keepaspectratio]{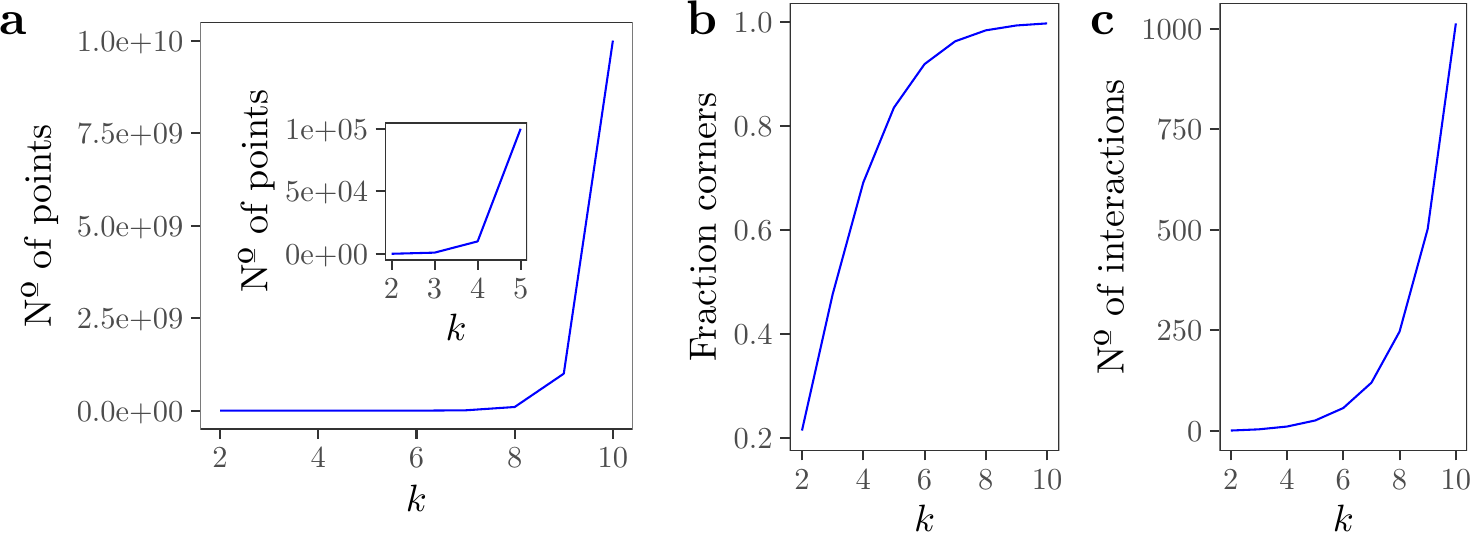}
\caption{The curse of dimensionality. a) Increase in the number of points needed to evenly sample an uncertainty space with a distance of 0.1 between adjacent points as a function of the number of model parameters $k$. b) Fraction of the space occupied by the corners of the hypercube along $k$ dimensions. c) Number of possible interactions between parameters as a function of $k$.}
\label{fig:dimension}
\end{figure}

Why do more detailed mathematical models tend to produce more uncertain estimates? To understand this paradox we draw attention to the concept of the model's \textit{uncertainty space}, which is the space formed by the number of uncertain parameters $k$. When $k=1$ the uncertainty space can be represented as a segment, when $k=2$ as a plane, when $k=3$ as a cube and when $k>3$ as a $k$-dimensional hypercube. Now let us assume that all parameters are distributed in the interval $[0, 1]$. Ten points suffice to sample the segment with a distance of 0.1 between points; if we want to keep the same sampling density to sample the plane, the cube and the $k$-dimensional hypercube, we will need 100, 1,000 and $10^k$ points respectively. Note how the number of sampling points increases exponentially with every addition of a new parameter (Fig.~\ref{fig:dimension}a). 

\begin{figure}[ht]
\centering
\includegraphics[keepaspectratio, scale = 1.4]{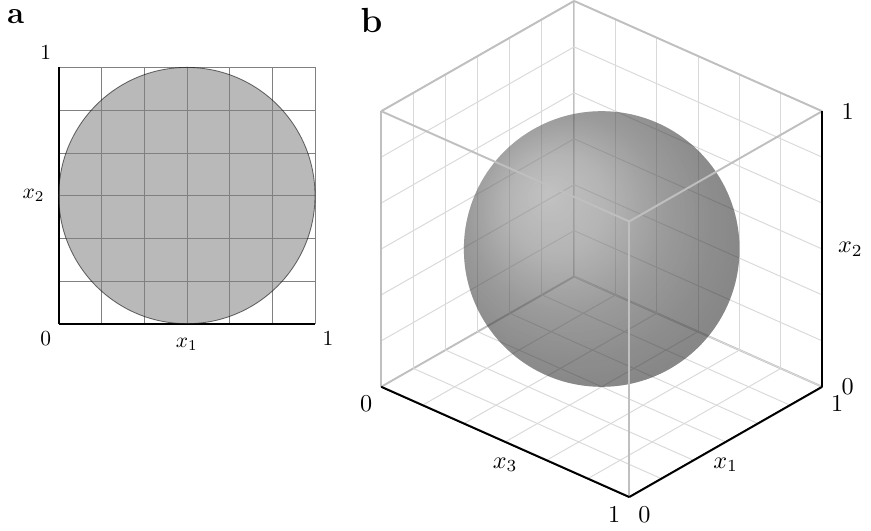}
\caption{Increase in the corners of the uncertainty space as a function of the number of parameters $k$. a) $k=2$. b) $k=3$.}
\label{fig:corners}
\end{figure}

The more parameters we add in our model, the larger the proportion of the uncertainty space (and hence the number of sampling points) that will be located in the corners and edges of the hypercube. To illustrate this fact, let us think about the ratio of the volume of an hypersphere of radius 1/2 to the volume of a unit hypercube of dimension $k$~\parencite{Saltelli2010b}. The center of the uncertainty space would equal the hypersphere whereas the corners would equal the space of the hypercube outside the hypersphere. For $k=2$ and $k=3$, these geometrical forms can be pictured as a circle inscribed in a plane and as a sphere within a cube. Consider how the volume of the space occupied by the corners increases exponentially with the addition of parameters: for $k=2$ it amounts to 21\% of the space (Fig.~\ref{fig:corners}a); for $k=3$ it corresponds to 47\% of the space (Fig.~\ref{fig:corners}b). By the time we reach $k=10$, the corners already form 99.7\% of the volume of the hypercube (Fig.~\ref{fig:dimension}b).

It is in the corners and edges of the hypercube where high-order effects, e.g., interactions between parameters whose effect on the model output uncertainty is larger than their individual effects~\parencite{Saltelli2008}, tend to occur. Non-additive operations such as multiplications, divisions or exponentials are enough to promote interactions, whose number grows as $2^k-k-1$. For a model with three parameters $f(\bm{x}), \bm{x}=(x_1,x_2,x_3)$, there might be four interactions up to the third-order, e.g., three two-order interactions ($x_1x_2$, $x_1x_3$, $x_2x_3$) plus one three-order interaction ($x_1x_2x_3$). The addition of one and two extra parameters rises the number of possible interactions to 11 and 26 and up to four and five-order effects respectively. With ten parameters, the number of possible interactions up to the tenth-order effect rises to 1,024 (Fig.~\ref{fig:dimension}c).

This explosion of the uncertainty space with every addition of a parameter means that finer-grained models have uncertainty spaces with disproportionally larger corners and potential higher-order interactions, a consequence of the curse of dimensionality~\parencite{Bellman1957}. If active, these high-order effects tend to boost the model output uncertainty. And the order of the highest-order effect active in the model tends to be higher in models with a larger number of uncertain parameters and structures~\parencite{Puyn}. This explains why the addition of model detail may not necessarily increase accuracy, but swamp the model output under indeterminacy. This becomes apparent \textit{if and only if} the model's uncertainty space is thoroughly explored. If it is not, its estimates will be spuriously accurate. And highly-detailed models tend to be so computationally demanding to run that the exploration of their uncertainty space often becomes unaffordable. 

\section*{Blackboxing}
\begin{figure}[ht]
\centering
\includegraphics[keepaspectratio]{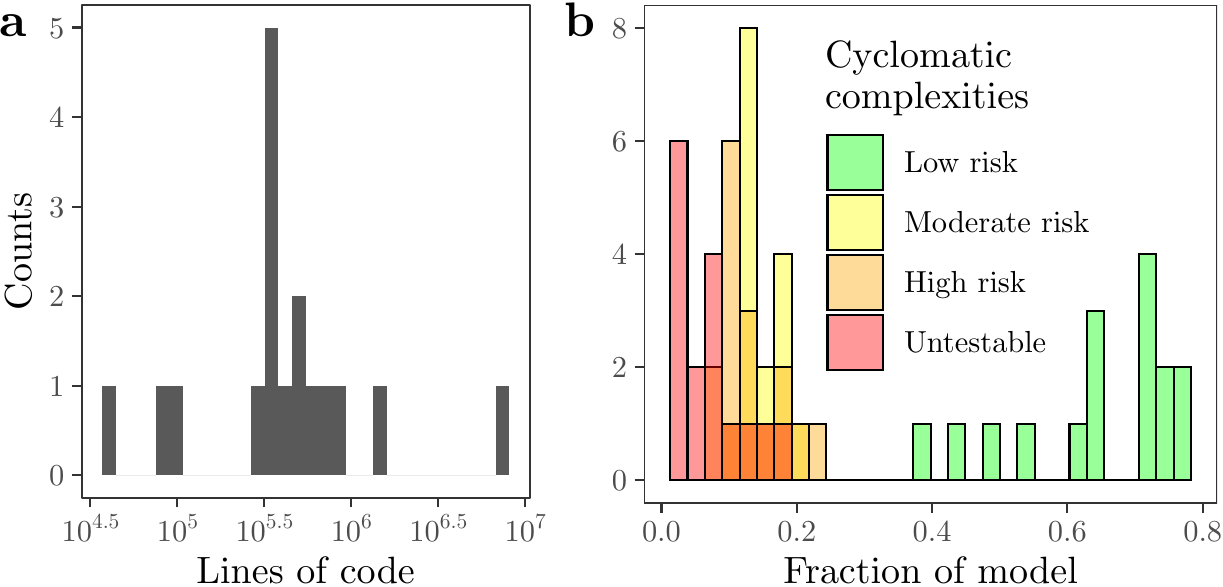}
\caption{Code of Global Climate Models (GCM). a) Lines of source code. b) Fraction of subprograms in each GCM with cyclomatic complexities, defined as the number of independent paths through a  program unit~\parencite{Mccabe1976}. Softwares with a higher cyclomatic complexity are harder to test and riskier to modify without introducing faults in the code, and are more prone to include bugs~\parencite{Chowdhury2010}. The plots are based on the data provided by~\textcite{Mendez2014} for 16 GCM.}
\label{fig:cyclomatic}
\end{figure}

The accumulation of parameters, feedbacks and interlinked compartments creates models whose behavior may not be easy to grasp intuitively. It is customary for models in the climate, environmental or epidemiological sciences to be formed by thousands of lines of code added up over decades and prone to include bugs, undesired behaviours and deprecated or obsolete language features. Global climate models, for instance, are on average composed of 500,000 lines of code in Fortran (with the model CESM of the Department of Energy of the National Science Foundation peaking at c.~1,300,000 lines of code) (Fig.~\ref{fig:cyclomatic}a), and include both obsolete statements and important cyclomatic complexities (Fig.~\ref{fig:cyclomatic}b)~\parencite{Mendez2014}. The model of the Imperial College London underpinning the response of the UK against covid-19 had c.~15,000 lines of code written over $\sim$13 years with several dangerous coding constructs, undefined behaviors and violations of code conventions~\parencite{Zapletal2021}. If we consider that in software development industry [whose quality control practices are more stringent than those of academia~\parencite{Kallen2021}] there may be on average 1-25 errors per 1,000 lines of code~\parencite{McConnell2004}, we may get to appreciate the number of undetected errors potentially hiding in the code of big scientific models. Lubarsky's Law of Cybernetic Entomology (``There is always one more bug'') looms larger in the quest for ever-exhaustive simulations.

Even under an optimistic scenario (e.g., only 10\% of the bugs are capable to meaningfully affect the results without the analyst noticing; only 50\% of the model is executed; 10 errors per 1,000 lines of code), the chance of obtaining wrong results with large models, say with a model formed by 20,000 lines of code, may become certainty~\parencite{Soergel2015}.

To increase quality and open up the model's black box several authors have requested modeling teams to give open source software licenses to their code and post it in repositories for inspection, error detection and reuse~\parencite{Morin2012, Barton2020}. The addition of comments and a user's manual may also improve usability and help pinpointing undesirable behaviors due to programming mistakes. These initiatives address the technical problems that arise when writing and upgrading code to improve the model's descriptive capacity, but fall short in handling another consequence of model hubris: the expansion in the number of value-laden assumptions embedded in the code~\parencite{Saltelli2020a}. Such features escape the formal checks used to locate and correct code defects and yet their impact in the simulations can easily be substantial: a bug in the code may offset the output of an algorithm that informs hurricane forecasts by 30\%~\parencite{Perkel2022}; the assumption that farmers prioritize long rather than short-maturing maize varieties can lead an algorithm to leave several million people without water insurance payouts~\parencite{Johnson2020}. 

A paradigmatic example of how the expansion of a model comes along with the addition of potentially problematic assumptions can be found in the use of Impact Assessment Models (IAMs) to guide policies against climate change. When the 1.5º~C goal became the target of climate action after the 2015 Paris agreement, IAMs were upgraded with modules that allowed the achievement of this target by assuming a wide and intensive use of negative emission technologies (e.g., bioenergy, carbon capture and storage)~\parencite{VanBeek2022, Anderson2016}. Such vision implied that a technology still in its infancy is deployed to fight climate change at a global scale and delivers as expected 15 years later. Many other questionable assumptions are likely to hide behind IAMs –-in fact, their number may be as large as to be unmanageable in the context of scientific publishing practices~\parencite{Skea2021}. This effectively places IAMs beyond the reach of peer-review~\parencite{Rosen2015}.

\section*{Computational exhaustion}
Until 2004--2006, the development of increasingly finer-grained models was fueled by computational advances that allowed to double the number of transistors in a chip while keeping the power requirements per unit of area constant. In other words: the speed of arithmetic operations could grow exponentially without significantly increasing power consumption. The doubling of the number of transistors and the scaling between chip area and power use are known as Moore's law and Dennard's scaling respectively~\parencite{Moore1965, Dennard1974}, and are two key trends to understand the evolution of computing performance and mathematical modeling over the last 50 years. Among others, they facilitated the onset of our current numerical, physics-based approach to model planetary climate change by sustaining the development of a ``hierarchy of models of increasing complexity'', as put by computeer pioneer and meteorologist Jule G. Charney~\parencite{Balaji2021}.

\begin{figure}[ht]
\centering
\includegraphics[keepaspectratio]{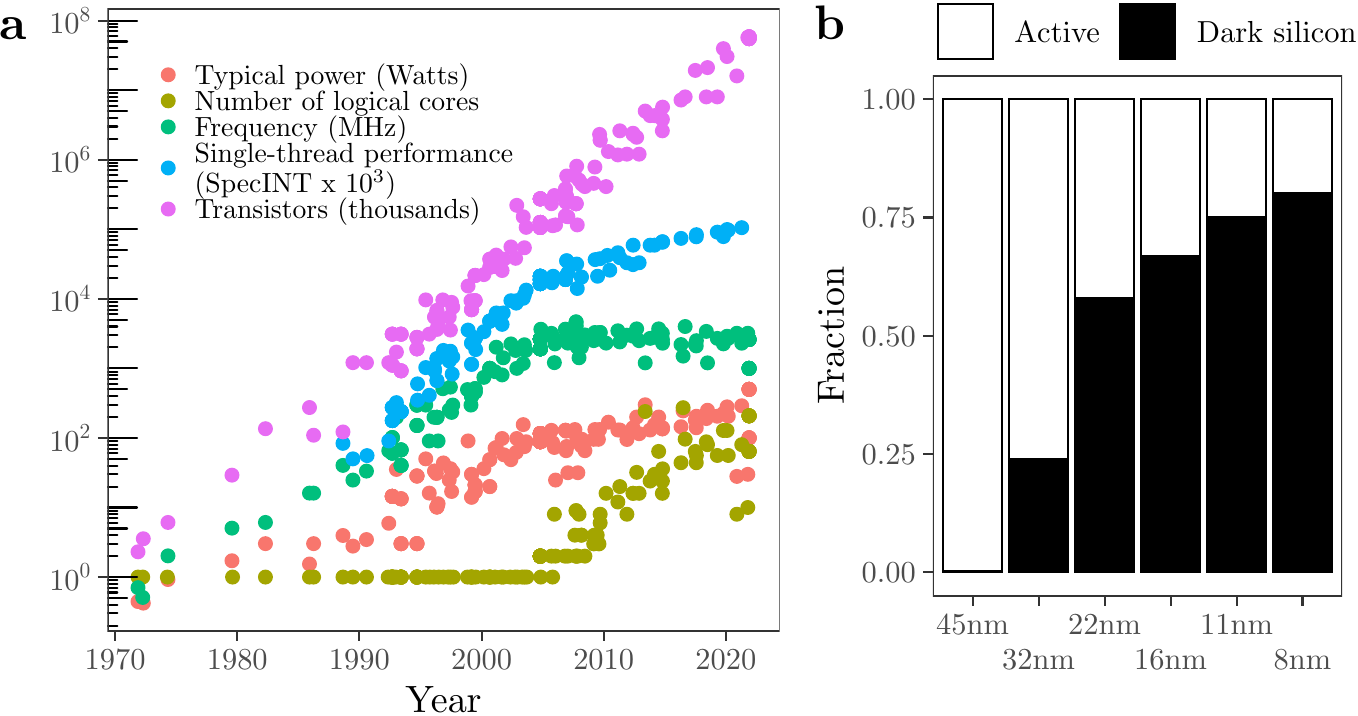}
\caption{Computational limits. a) Microprocessor trend data up to the year 2020, retrieved from Karl Rupp in \url{https://github.com/karlrupp/microprocessor-trend-data} (Accessed 25th of May 2022). b) Fraction occupied by ``dark silicon'' as a function of the technology ($x$ axis). Modified from Fig.~1.1 in~\textcite{Kanduri2017}.}
\label{fig:computation}
\end{figure}

After 2006 it became apparent that this trend was over (Fig.~\ref{fig:computation}). Basically, the doubling in the number of transistors in a chip could no longer be matched by a proportional boost in energy efficiency. This resulted in chips with increased power density and more heat generated per unit of area, which had to be dissipated to prevent thermal runaway and breakdown. A solution was to keep a fraction of the chip power-gated or idled --the part known as ``dark silicon''~\parencite{Taylor2012}. The higher the density in the number of transistors, the larger the fraction of the chip that has to be underused, a limitation that tapers off the improvement in clock frequencies and single-thread performances (Fig.~\ref{fig:computation}a). Since higher chip densities are being required to match the demands of machine learning, big data and mathematical models, the fraction occupied by ``dark silicon'' may soon be larger than 90\%~\parencite{Taylor2012, Kanduri2017} (Fig.~\ref{fig:computation}b). Adding extra cores to improve performance is unlikely to sort this issue out due to its meagre computational returns~\parencite{Esmaeilzadeh2011}.

Power consumption hence imposes a severe constraint to the addition of model resolution and suggests that, under our current computational paradigm, the quest towards ever-detailed models is computationally unsustainable. And the problem is not only one of arithmetic performance, but also of data storage. The 100 models participating in phase 6 of the Coupled Model Intercomparison Project currently produce c. 80 petabytes of data [1 petabyte (PB) = 1,000 terabytes (TB)]; if they increase resolution at a $\sim$1km, the output may reach 45,000 PB~\parencite{Schar2020}, $\sim$2,200 times more data than that stored in the Library of Congress in 2019 ($\sim20$ PB)~\parencite{Spurlock2019}. To tackle the data avalanche derived from going hyperresolution some authors have suggested to just save the simulation setup and re-run the simulation on demand~\parencite{Schar2020}, yet this strategy may breach the FAIR data principles whereby data should be findable, accessible, interoperable and replicable~\parencite{Wilkinson2016}, and re-computing may be even more costly than archiving~\parencite{Bauer2015}.

\section*{Model attachment}
The pursuit of finer-grained models requires the training of specialists able to set, calibrate, upgrade and analyze these models and pull appropriate insights for scientific and policy purposes. The acquisition of this expertise is time-consuming and usually demands several years of education at a university or research institution. As with any other learning process, this situation sets the ground for the development of a domain-specific knowledge that endows its possessor with the capacity to efficiently put the learned skills to use in exchange for a higher risk of suffering cognitive biases such as~\textcite{Maslow1966}'s hammer  --``If the only tool you have is a hammer, it is tempting to treat everything as if it were a nail''. In mathematical modeling this phenomenon takes place when a model is used repeatedly regardless of purpose or adequacy (a model of ``everything and everywhere''), a bias that has been attested for hydrological modeling: based on a sample size of c.~1,500 papers and seven global hydrological models (GHM),~\textcite{Addor2019} observed that in 74\% of the studies the model selected could be predicted based solely on the institutional affiliation of the first author. 

\begin{figure}[ht]
\centering
\includegraphics[keepaspectratio]{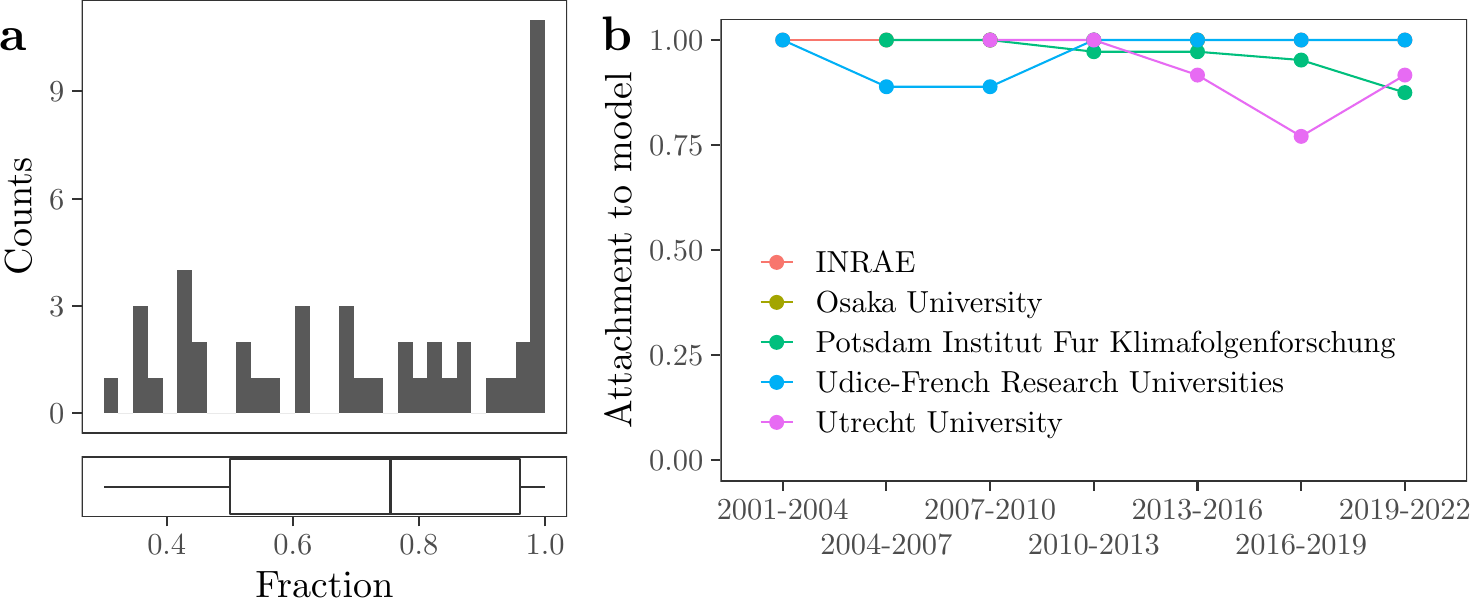}
\caption{Institutional attachment to global models. a) Strength of the institutional attachment to a favorite model. A value of 1 means that 100\% of the studies published by a given institution rely on the same global model. Only institutions with more than 5 articles published are considered. b) Strength of the attachment for some specific institutions through time.}
\label{fig:summary_legacy}
\end{figure}

\begin{figure}[!ht]
\centering
\includegraphics[keepaspectratio]{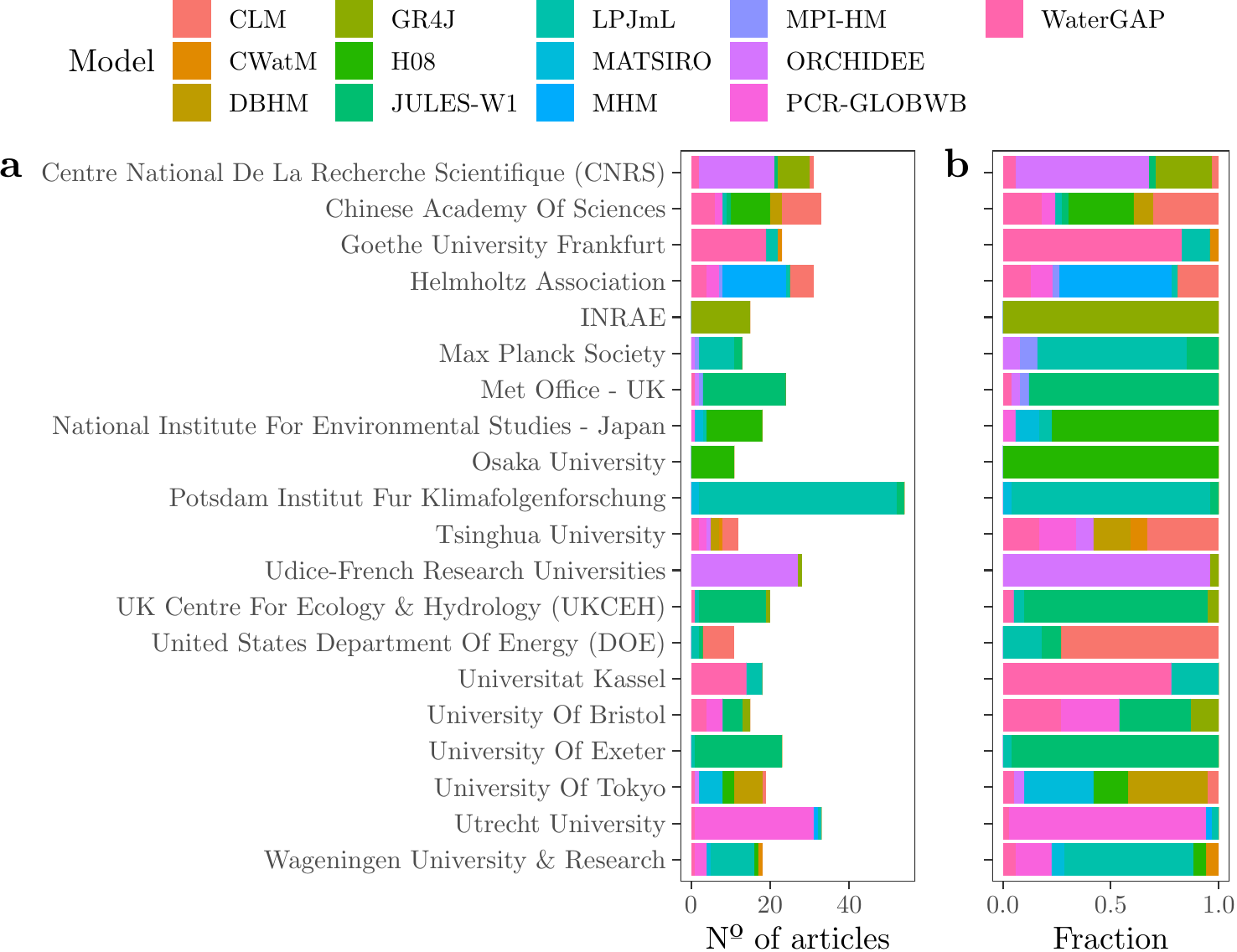}
\caption{Institutional attachment to global models. Only the 20 most productive institutions are displayed. a) Stacked barplot with the number of publications per institution. b) Fractioned stacked barplot.}
\label{fig:legacy}
\end{figure}

Here we extend the work by~\textcite{Addor2019} to 13 new global (hydrological, vegetation and land surface) models across c.~1,000 papers [CLM, CWatM, DBHM, GR4J, H08, JULES-W1, LPJmL, MATSIRO, MHM, MPI-HM, ORCHIDEE, PCR-GLOBWBW and WaterGAP; see~\textcite{Puy2022_hubriscode} for the methods and the references], with the same results: the institutional affiliation of the first author anticipates the model used in $\sim75$\% of the publications (Fig.~\ref{fig:summary_legacy}a). In fact, several institutions display a very high level of attachment. This is the case of the INRAE (100\%, GR4J, 15 papers), Osaka University (100\%, H08, 11 papers), Utrecht University (91\%, PCR-GLOBWB, 33 papers), the Potsdam Institut Fur Klimafolgenforschung (93\% LPJmL, 54 papers), the University of Exeter (96\%, JULES-W1, 23 papers) or Udice --Universités de Recherche Françaises (96\%, ORCHIDEE, 28 papers) (Fig.~\ref{fig:legacy}a). The attachment of these institutions to their favourite model is also very consistent through time (Fig.~\ref{fig:summary_legacy}b).

The case of global models illustrates the extent to which model selection may be influenced by path-dependencies that do not necessarily match criteria of adequacy, but rather convenience, experience and habit~\parencite{Addor2019}. This inertia might be seen as an instance of Einstellung effect, the insistence on using familiar tools and frames to tackle a problem even if they are sub-optimal or inadequate~\parencite{Luchins1942}. Very large models such as global models may become institutionalized to offset in the mid-term the costs derived from setting up a more efficiency ecosystem in terms of writing of code, training of personnel, selection of modeling methods, calibration algorithms, upgrading, etc. Once a given modeling ecosystem is in place, the workflow is streamlined and the researcher has ``to do nothing but to press a few enters and then [the modeling pre-processement] is done''~\parencite[15]{Melsen2022}. Switching to other models resets this process and moves the watch back for the institution and the researcher. Larger models may be more prone to promote model attachment and become models of ``everything and everywhere'' given the much higher costs involved in the first stages of their institutionalization.

\section*{Concluding remarks}
In this chapter we have suggested moderating our Cartesian thirst to unfold the secrets of the natural world via increasingly detailed mathematical models. This pursuit may lead to overlook important discussions over mathematical Platonism and the nature of mathematical knowledge. It may also promote black-boxing, computational exhaustion and model attachment. There exists in mathematical modeling a ``political economy'' whereby larger models command more epistemic authority and offer to their developer(s) better opportunities to defend them against criticisms. Thus the resistance of some modelers to come to terms with the full uncertainty of their work may have motivations such as that of ``navigating the political''~\parencite{VanBeek2022}, i.e., defending the role of modeling work in policy-relevant settings, where epistemic authority needs to be preserved~\parencite{Robertson2021}.

Modeling is but one among many instances of quantification where the issue of overambition, in the words of~\textcite[36]{Quade1980}, is a recurring theme. In that sense, sociology of quantification offers powerful instruments to dissect the normative implications of our ambitions to box reality behind numbers. There are instances of quantification where contestation comes more natural, such as in the use of statistical indicators in socio-economical domains, discouted by \textit{statactivists}~\parencite{Bruno2014, Mennicken2022}. Even when it comes to rating and ranking there is a rich literature condemning the excesses of quantification~\parencite{Muller2018, ONeil2016}, and a recent success of activism in this field is the fight against the Work Bank’s Doing Business index~\parencite{Cobham2022}. The use of statistical rituals in sociology is observed by~\textcite{Gigerenzer2015} and the mathematization of the economy is discussed by Lukas and Drechsler in another chapter of this book. As argued elsewhere~\parencite{Saltelli2019b}, mathematical modelling tends to be shielded from more stringent forms of critical activism by the barrier that the complexity of the mathematical constructs and coding pose to experts and non-experts alike. 

\section*{Data availability}
The code to reproduce the results and the figures of this study can be found in~\textcite{Puy2022_hubriscode} and in GitHub (\url{https://github.com/arnaldpuy/model_hubris}).

\printbibliography

\end{document}